\documentclass{aastex}
\usepackage{spr-astr-addons}
\usepackage{url}

\RequirePackage{color}

\begin{document}

\title{On the spin distributions of $\Lambda$CDM haloes}
\shorttitle{On the spin distributions of $\Lambda$CDM haloes}
\shortauthors{N. Hiotelis}

\author{N. Hiotelis \altaffilmark{1}}
\affil{1st Experimental Lyceum of Athens, Ipitou 15, Plaka, 10557,
Athens,  Greece} \email{hiotelis@ipta.demokritos.gr}
\altaffiltext{1}{Lysimahias 66, Neos Kosmos, Athens, 11744 Greece,
e-mail:hiotelis@ipta.demokritos.gr }

\begin{abstract}
We used merger trees realizations, predicted by the extended
Press-Schechter
      theory, in order to study  the growth of angular momentum of dark matter haloes.
      Our results showed that:\\
      1) The spin parameter  $\lambda'$  resulting from the above method, is an increasing function of the present
      day mass of the halo.  The mean value of $\lambda'$ varies from 0.0343 to 0.0484 for haloes with present
      day masses in the range of $ 10^9\mathrm{h}^{-1}M _{\odot}$ to
      $10^{14}\mathrm{h}^{-1}M_{\odot}$.\\
      2)The distribution of $\lambda'$ is close to a  log-normal , but, as it is already found in
      the results of N-body simulations, the match is not satisfactory at the tails of the distribution.
      A new analytical formula that approximates the results much more satisfactorily is presented.\\
      3) The distribution of the values of $\lambda'$ depends only weakly on the redshift. \\
      4) The spin parameter of an halo depends on the number of recent major mergers. Specifically
      the spin parameter is an increasing function of this
      number.
\end{abstract}

%


\section{Introduction}

There are two more likely pictures regarding the growth
of angular momentum in dark matter  haloes.\\
 The first is that galactic haloes acquired their angular momentum from the tidal torques of
the surrounding matter. This is an old idea of Hoyle (1949) that
has been investigated in a large number of studies (e.g. Peebles
1969; Doroshkevich 1970;  Efstathiou \& Jones 1979, Barnes \&
Efstathiou 1987, White 1984; Voglis \& Hiotelis 1989, Warren et
al. 1992, Steinmetz \& Bartelmann 1995). The results of the above
studies, analytical and numerical, show that the spin parameter
$\lambda$, introduced by Peebles (1969) and defined by the
relation $\lambda\equiv J \sqrt{\mid E\mid}/GM^{5/2}$, has an
average value of about 0.05-0.07, where $J$ is the modulus of the
spin of halo, $M,E $ are the mass and the energy respectively and
$G$ is the gravitational constant. According to the above picture,
haloes acquire their angular momentum during their linear stage of
their evolution because during this stage they have large linear
sizes and thus the environment is capable to affect their
evolution by tidal torques. Since their expansion is decelerating
their relative linear size becomes smaller and the affection by
the environment becoms less significant. The moment of the maximum
expansion is in practice the end of the epoch of growth of angular
momentum. At latter times, the halo evolves as a dynamically
isolated system. Steinmetz \& Bartelmann  (1995) showed that the
dependence of the probability distribution of $\lambda$ on the
density parameter of the model Universe as well as on the variance
of the density contrast field is very weak. Only a marginal
tendency for $\lambda$ is found to be larger for late-forming
objects in an
open Universe.\\
The second picture is closely related to the hierarchical
clustering scenario of cold dark matter (CDM; Blumenthal et al.
1986). According to this scenario, structures in the Universe grow
from small initially Gaussian density perturbations that
progressively detach from the general expansion, reach a maximum
radius and then collapse to form bound objects. Larger haloes are
formed hierarchically by mergers between smaller ones, called
progenitors. The buildup of angular momentum is a random walk
process associated with the mass assembly history of the halo's
major progenitor. The main role of tidal torques in this picture
is to produce the random tangential velocities of merging
progenitors.\\
The above two pictures of formation are usually studied by two
different kinds of methods. The first kind is the N-body
simulations that are able to follow the evolution of a large
number of particles under the influence of the mutual gravity,
from initial conditions to the present epoch. The second kind
consists of semi-analytical methods. Among them, the
Press-Schechter (PS) approach and its extensions (EPS) are of
great interest since they allow to compute  mass functions (Press
\& Schechter  1974; Bond et al. 1991) to approximate merging
histories (Lacey \& Cole 1993,  Bower 1991, Sheth \& Lemson 1999b)
and to estimate the spatial clustering of dark matter haloes (Mo
\& White 1996; Catelan et al.
1998, Sheth \& Lemson 1999a).\\
Recently large cosmological N-body simulations have been performed
in order to study the angular momentum of dark matter haloes  in
$\Lambda CDM$ models of the Universe (e.g. Bullock 2001, Kasun \&
Evrard 2005, Bailin \& Steinmetz 2005, Avila-Reese
et al. 2005, Gottl\"{o}ber  \& Turchaninov 2006).\\
Additionally, semi-analytical methods like merging histories
resulting from EPS methods, have been used for similar studies
(Vitvitska et. al 2002, Maller et. al 2002). \\
In this paper, we use such merging histories based on EPS
approximations to study the distribution of spins.\\
The paper is organized as follows: In Sect.2, basic equations are
summarized. In Sect.3, we present our results while a discussion
is given in Sect.4.
\begin{figure}[t]
  \includegraphics[width=8cm]{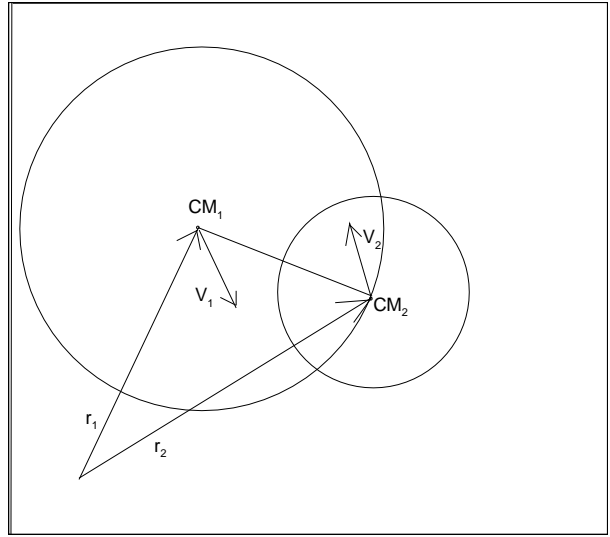}
    \caption{The figure shows  the conditions at the onset of the merger of two haloes with masses $m_1$ and $m_2$ and virial radii
    $R_1$ and $R_2$
    respectively with $m_1 > m_2$ and $R_1 > R_2$. The centers of their masses are indicated by $CM_1$ and $CM_2$ and their distance
    is $r\equiv max(R_1,R_2)=R_1$. The position vectors of their centers of mass are $\mathbf{r_1}$
    and $\mathbf{r_2}$ respectively so their  relative position is
    $\mathbf{CM_{12}}=\mathbf{r_2}-\mathbf{r_1}$.
    The vectors of the velocities of their center of mass are $\mathbf{v_1}$ and $\mathbf{v_2}$ and the vector
     of their relative velocity is
     $\mathbf{v_{rel}}=\mathbf{v_2}-\mathbf{v_1}$.
        Haloes merge if they approach each other, that is the condition
         $\mathbf{v_{rel}}\cdot \mathbf{CM_{12}} <0$ is fulfilled.
         See text for more details.
   }\label{fg1-eps}
    \end{figure}
      \begin{figure}[t]
  \includegraphics[width=8cm]{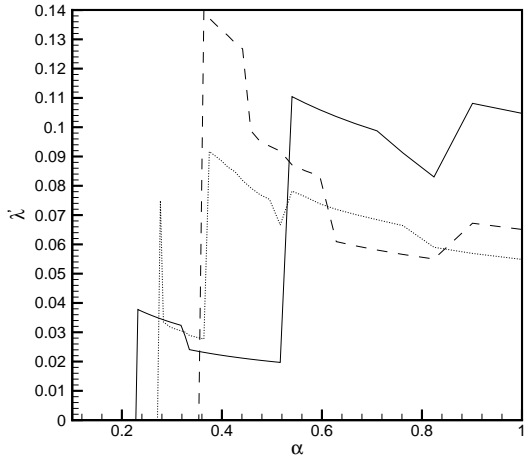}
    \caption{Spin parameter  $\lambda'$ for the most massive progenitor at scale factor $\alpha$ for the cases $1, 2$ and
  $3$. Solid, dashed and dotted lines correspond to the cases 1, 2 and 3 respectively. The evolution of $\lambda'$ is characterized by sharp increases and decreases due to
  mergers.}\label{fg2-eps}
    \end{figure}
\section{Construction of merger-trees and acquisition of angular momentum}
  According to the hierarchical scenarios of structure formation, a
  region collapses at time $t$ if its overdensity at that time
  exceeds  some threshold. The linear extrapolation of this
  threshold up to the present time is called a barrier, B. A
  likely form of this barrier is:
  \begin{equation}
  B(S,t)=\sqrt{aS_*}[1+\beta(S/aS_*)^{\gamma}].
  \end{equation}
  In the above Eq. $\alpha$, $\beta$ and $\gamma$ are constants,
   $S_*\equiv S_*(t)\equiv \delta^2_c(t)$ where $\delta_c(t)$ is the linear extrapolation
  up to the present day of the initial overdensity of  a spherically symmetric  region that collapsed at
  time $t$. Additionally, $S\equiv \sigma^2(M)$, where $\sigma^2(M)$ is the present day  mass
  dispersion on  comoving  scale containing mass $M$. $S$ depends on the assumed
  power spectrum.
  The spherical collapse model  has a barrier that does not
  depend on the mass (eg. Lacey \& Cole 1993). For this model, the  values of the parameters are
  $a=1$ and $\beta=0$. The ellipsoidal  collapse model (EC)
  (Sheth, Mo  \& Tormen 2001) has
   a barrier that depends on the mass (moving barrier). The values of the
   parameters are $a=0.707$,$\beta=0.485$,  $\gamma=0.615$ and are adopted
    either from the dynamics of ellipsoidal collapse or from
  fits to the results of N-body simulations.  \\
   Sheth \& Tormen  (2002) showed
   that given  a mass element -that is a
  part of a halo of mass $M_0$ at time $t_0$-  the probability that at earlier
  time $t$ this mass element was a part of a smaller halo with mass $M$, is
  given by the equation:
  \begin{equation}
  f(S,t/S_0,t_0)\mathrm{d}S=\frac{1}{\sqrt{2\pi}}\frac{|T(S,t/S_0,t_0)|}{(\Delta
  S)^{3/2}}
  \exp\left[-\frac{(\Delta B)^2}{2\Delta S}\right]\mathrm{d}S
  \end{equation}
  where $\Delta S=S-S_0 $ and $\Delta B=B(S,t)-B(S_0,t_0)$ with
  $S=S(M), S_0=S(M_0)$.\\
  The function $T$ is given by:
  \begin{equation}
  T(S,t/S_0,t_0)=B(S,t)-B(S_0,t_0)+\sum_{n=1}^{5}\frac{[S_0-S]^n}{n!}
  \frac{\mathrm{\partial ^n}}{\partial S^n}B(S,t).
  \end{equation}
  Recently, Zhang \& Hui (2006) derived first crossing  distributions of random walks
  with a moving barrier of an arbitrary shape. They showed that this distribution satisfies an
  integral equation that can be solved by a simple matrix inversion. They compared the predictions
  of their exact numerical solution with those of approximation given by eq.3 and found a very good agreement. This
  shows that eq. 3 works well for mildly non linear barriers as that given by eq.1 above.\\
   Eq. 2. can obviously predict the unconditional mass probability, $f(S,t)$,
  which is simply the probability that a mass element is at time
  $t$ a part of a halo of mass $M$, by setting $S_0=0$, and
  $B(S_0,t_0)=0$. We note that the quantity $Sf(S,t)$
  is a function of the variable $\nu$ alone, where $\nu\equiv
  \delta_c(t)/\sigma(M)$. $\delta_c$ and $\sigma$ evolve
  with time in the same way, thus the quantity $Sf(S,t)$ is independent of
  time. Setting $2Sf(S,t)=\nu f(\nu)$ one obtains the so-called
  multiplicity function $f(\nu)$. The multiplicity function is the distribution
   of first crossings of  a barrier $B(\nu)$ (that is why the shape of the barrier
   influences the form of the multiplicity function), by independent uncorrelated Brownian
    random walks (Bond et al. 1991). The multiplicity function
   is related to the comovimg number density of haloes of mass $M$ at time $t$, $N(M,t)$,
   by the relation,
  \begin{equation}
  \nu f(\nu)=\frac{M^2}{\rho_b(t)}N(M,t)\frac{\mathrm{d}\ln M}{\mathrm{d}\ln \nu}
  \end{equation}
  that results from the excursion set approach (Bond et al.
  1991). In the above Eq., $\rho_b(t)$ is the density of the model of the Universe at time
  $t$.\\
  Using a barrier of the form of Eq.1 in the unconditional mass probability, one finds for
  $f(\nu)$ the expression:
  \begin{equation}
  f(\nu)=\sqrt{2a / \pi}[1+\beta(a
  {\nu}^2)^{-\gamma}g(\gamma)]\exp\left(-0.5a\nu^2[1+\beta(a\nu^2)^{-\gamma}]^2\right)
  \end{equation}
  where
  \begin{equation}
  g(\gamma)=
  \mid 1-\gamma +\frac{\gamma (\gamma
  -1)}{2!}-...-\frac{\gamma(\gamma-1)\cdot \cdot \cdot
  (\gamma-4)}{5!} \mid
  \end{equation}
  Recent comparisons  show that the use of EC model improves the agreement
between the results of EPS methods and those of N-body
simulations. For example,  Yahagi et al. (2004) show that the
multiplicity function resulting from  N-body simulations is far
from the predictions of spherical model while it shows an
excellent agreement with the results of the EC model. On the other
hand, Lin et al. (2003) compared the distribution of formation
times of haloes formed in N-body simulations with the formation
times of haloes formed in terms of the spherical collapse model of
the EPS theory. They found that N-body simulations give smaller
formation times (i.e.higher redshifts).  Hiotelis \& Del Popolo
(2006) showed that using the EC model, formation times are shifted
to smaller values than those predicted by a spherical collapse
model. Additionally, EC model, combined with the 'stable
clustering' hypothesis, was used in order to study density
profiles of dark matter haloes (Hiotelis 2006). The resulting
density profiles at the central regions of haloes have the
interesting feature to be closer to the results of observations
than the results of N-body simulations. Consequently, the EC model
is a significant improvement of the spherical model and therefore
we use this model for our calculations.\\
     We assume a number of haloes with the same present day mass $M_0$ -at present epoch $t_0$- and we study their
  past using merger-trees by finding their progenitors -haloes that merged
  and formed the present day haloes- at previous times. The procedure for a single halo is
  as follows:  A new time $t<t_0$ is chosen. Then, a value $\Delta S$ is chosen from the desired distribution
  given by Eq.2. The mass $M_p$ of a progenitor is found by solving for $M_p$ the equation
  $\Delta S=S(M_p)-S(M_0)$. If the mass left to be resolved $M_0-M_p$ is large enough, the above procedure is repeated
  so a distribution of the progenitors of the halo is created at $t$. If the mass left to be resolved -that equals to
  $M_0$ minus the sum of the masses of its progenitors- is less than a threshold then, we proceed to the next
  time analyzing with the same procedure the mass of each progenitor. The most massive progenitor at
  $t$ is considered as the mass of the initial halo at that time. \\
     A complete description of the above numerical method is given in
  Hiotelis \& Del Popolo (2006). The algorithm - known as N-branch merger-tree-
  is based on the pioneered works of  Lace \& Cole (1993),
  Somerville \&  Kollat (1999) and van den  Bosch (2002).\\
     In our calculations, we used a flat model for the Universe with
  present day density parameters $\Omega_{m,0}=0.3$ and
   $ \Omega_{\Lambda,0}\equiv \Lambda/3H_0^2=0.7$.
  $\Lambda$ is the cosmological constant and $H_0$ is the present day value of Hubble's
  constant. We used the value $H_0=100\mathrm{hKMs^{-1}Mpc^{-1}}$
  and a system of units with $m_{unit}=10^{12}M_{\odot}h^{-1}$,
  $r_{unit}=1h^{-1}\mathrm{Mpc}$ and a gravitational constant $ G=1$. At this system of
  units,
  $H_0/H_{unit}=1.5276.$\\
  As regards the power spectrum,  we  used the $\Lambda CDM$ form proposed by
  Smith et al. (1998). The power spectrum is smoothed using the top-hat window function and
  is normalized for $\sigma_8\equiv\sigma(R=8h^{-1}\mathrm{Mpc})=1$.\\
  A merger-tree gives the complete history of the halo. After its
  construction,  we know all the progenitors of a halo with present-day, at time $t=t_0$,  mass $M_0$ at a previous time $t_1$,
  all progenitors of every progenitor at $t_1$ at time $t_2 < t_1$
  etc. This procedure is repeated up to a level $n$ of resolution that corresponds to a time
  $t_n.$
  As regards the choice of the time-step we used the procedure described in Hiotelis \& Del Popolo
  2006 that is: The equation $\delta_c(a_{new})=\Delta \omega +\delta_c(a_{old})$, where $a_{old}=1$ at the beginning of the
  construction,
  is solved for $a_{new}$, that is the new value of the scale factor.
  We used a constant value of $\Delta\omega =0.1$ but tests with
  smaller values of $\Delta\omega$ showed no differences in the
  results. The procedure stops for $a<a_{min}=0.05$.
   Then, we start to merge haloes present at time $t_n$ to create the angular momentum history of those haloes
  that are present to the time level $t_{n-1}$. We assume that two haloes with virial masses $m_1$ and $m_2$ and
  virial radii  $r_1$ and $r_2$ merge when the following conditions are fulfilled:\\
  a) They approach each other.\\
  b) Their relative energy, given by $E=\frac{1}{2}\mu
  v^2-\frac{Gm_1m_2}{r}$,  is negative, where $\mu \equiv \frac{m_1m_2}{m_1+m_2}$, $r$ and $v$
  are the distance and the relative velocity of their canters of masses.\\
  c) Their distance $r$ is equal to the the maximum of $r_1$ and
  $r_2$.\\
  After such a merge a new halo is created with mass $m_{1,2}=m_1+m_2$  and spin $\mathbf{S_{1,2}}$ given by
  \begin{equation}
  \mathbf{S_{1,2} }=\mathbf{S_1}+\mathbf{S_2}+\mathbf{L_{orb,1,2}}
  \end{equation}
  where $\mathbf{S_1}$ and  $\mathbf{S_2}$ are the spins of the
  two haloes and $\mathbf{L_{orb,1,2}}$ is their orbital angular momentum given by
  \begin{equation}
   \mathbf{L_{orb,1,2}}=\frac{m_1m_2}{m_1+m_2}(\mathbf{r}\times\mathbf{v})
   \end{equation}
   $\mathbf{r}$ and $\mathbf{v}$ are the vectors of relative
   position and velocity of their center of mass respectively.
   A halo which has  suffered no merger up to a time $t$ has no spin and
   consequently all haloes at time $t_n$ have no spin.
   The virial $r$ radius of an halo at scale factor $a$ is related to its virial mass $m$ by the
   relation:
   \begin{equation}
    r(a)=\left[\frac{2Gm(a)}{X(a)}\right]^{1/3}
    \end{equation}
    where
    \begin{equation}
    X(a)\equiv \Delta_{vir}(a)\Omega_m(a)H^2(a).
    \end{equation}
    $\Omega _m(a)$ and $H(a)$ are the density parameter and the Hubble's constant at
    scale factor $a$, respectively. For $\Delta_{vir}$ we used the expression
    given in Bryan \& Norman (1998) $\Delta_{vir}(a)\approx
    (18\pi^2-82x-39x^2)/\Omega_m(a)$ where $x\equiv
    1-\Omega_m(a)$.\\
    The construction of a merger-tree does not requires or
    predicts any information about the velocity field of merging
    haloes. Since, in our case, the purpose is to study the growth of angular momentum during a process of subsequent
    mergers, we need a model for the description of the velocity field. So, at first, we used an arbitrary, but reasonable,
    model, that is described in details below,satisfying the conditions a) b) and c) set
    above.Additionally in section 3. we also refer to a model for the velocity
    field that is consistent with the results of N-body simulations. Both models gave similar results, but the
    question which model describes the velocity field best is open and under
    investigation.\\
    The whole procedure of merging the haloes present in the merger-tree model
    follows:\\
     Let that at level $l$, a set $k$  haloes with virial masses $m_{l,1},m_{l,2}...    m_{l,k}$, virial radii
    $r_{l,1},r_{l,2}...r_{l,k}$ and spins $\mathbf{S}_{l,1},\mathbf{S}_{l,2}...\mathbf{S}_{l,k}$  consists of all
    the progenitors of an halo with virial mass $m_{l-1,1}$
    and virial radius $r_{l-1,1}$ at level $l-1$. The merger procedure is as follows:
    Two progenitors $m_{l,1}$ and $m_{l,2}$ merge. First $r=max(r_{l,1},r_{l,2})$ and $\mu \equiv
    \frac{m_{l,1}m_{l,2}}{m_{l,1}+m_{l,2}}$ are calculated. Then, the maximum
    relative velocity, that satisfies the condition of negative total orbital energy, is found by the
    relation $v_{max}=(\frac{2Gm_{l,1}m_{l,2}}{r\mu})^{1/2}$.
    Then, the modulus, $v_{rel}$, of the relative velocity vector $\mathbf{v}_{rel}$ of the two haloes  is picked by a Gaussian distribution with mean value $v_{mean}=
    v_{max}/2$ and $\sigma =(1/3)v_{mean}$. The two components $v_x$ and $v_y$   are found using uniform distributions in the range $[-v_{rel},v_{rel}]$. If
    the condition $v^2_x+v^2_y \leq v^2_{rel}$ is satisfied, the third component if found by
    $v_z=\pm \sqrt{v^2_{rel}-(v^2_x+v^2_y)}$, where the sign is
    chosen randomly. If the above inequality is not satisfied, new
    values of $v_x$ and $v_y$ are chosen and the procedure is repeated. The components ($x,y,z$)of the relative
    position vector $\mathbf{r}$, with modulus $r\equiv max(r_{l,1},r_{l,2})$ are defined choosing $x$ and $y$ by a uniform distribution in
    the range $[-r,r]$ and then, if the condition $x^2+y^2 \leq r^2$ is fulfilled , the component $z$ is
    defined by by $z=\pm \sqrt{r^2-x^2-y^2}$.
     The condition of approaching, $\mathbf{r\cdot v}_{rad}\leq 0$, where
     $\mathbf{v}_{rad}$ is the radial component of the relative velocity, is checked. If the condition is not fulfilled
     we go back to choose new velocity components,  otherwise we continue by finding the
     orbital angular momentum, according to (8) and the spin of the newly formed halo according to (7).
     The mass of this halo is $m_{l,1}+m_{l,2}$, and its
    virial radius is defined by (9) where $a$ is the scale factor of the Universe at level
    $l$. The number of $k$ haloes of the set is now  reduced by one. The procedure is repeated until the number
    of haloes becomes one. Thus, the angular momentum of a new halo, present at time level $l-1$, which had at time level $l$ the above
    $k$ progenitors, is found. The procedure is repeated for all the sets of progenitors at level $l$ and so the
    angular momentum for every halo of level $l-1$ is  calculated. This new set of haloes consists of the progenitors of haloes at level $l-2$. The same procedure
    is repeated for level $l-1$ to  create the haloes at level $l-2$ and so on.  The procedure ends with the formation of a single
    halo of mass $m_0$ at level $0$ that represents the present age $t_0$ of the
    Universe. Fig.1 shows two haloes at the onset of their merger. \\
    As a measure of the angular momentum, we used the parameter ${\lambda}'$
    given by the (see Bullock et al. 2001; Dekel et al. 2000)
    \begin{equation}
    {\lambda}'\equiv \frac{S}{\sqrt{2}mv_{c}r}
    \end{equation}
    where $S$ equals to the spin, $m$ is the virial mass, $r$ the virial radius of the halo
    and $v_{c}=\sqrt{Gm/r}$ is the circular velocity at
    distance $r$. The parameter ${\lambda}'$ is easier to
    be measured, than ${\lambda}$,  not only in simulations but in semi-analytical methods as the merger-tree method used in this paper.
    This happens because the total energy of the halo is not
    required. Instead, the calculation of $\lambda$ requires a known density profile $\rho$ for the halo, the
    calculation of the potential $\phi$ by the expression
    \begin{equation}
    \Phi(r)=-4\pi
    G[\frac{1}{r}\int_0^r\rho(r'){r'}^2\mathrm{d}r'+\int_r^{\infty}\rho(r')r'\mathrm{d}r'],
    \end{equation}
    the calculation of the potential energy $W$
    \begin{equation}
    W=\frac{1}{2}\int \rho(r)\Phi(r)4\pi r^2\mathrm{d}r
    \end{equation}
    and finally, assuming virial equilibrium, the derivation of the total energy by $E=\frac{1}{2}W$.Simplified forms for
    specific density profiles can be found in Sheth et. al 2001\\
    It is noticed that the spin parameters $\lambda$ and $\lambda'$ are approximately equal for typical Navarro et al.
    1997 haloes  (Bullock et al. 2001).

\begin{figure*}[t]
  \includegraphics[width=17cm]{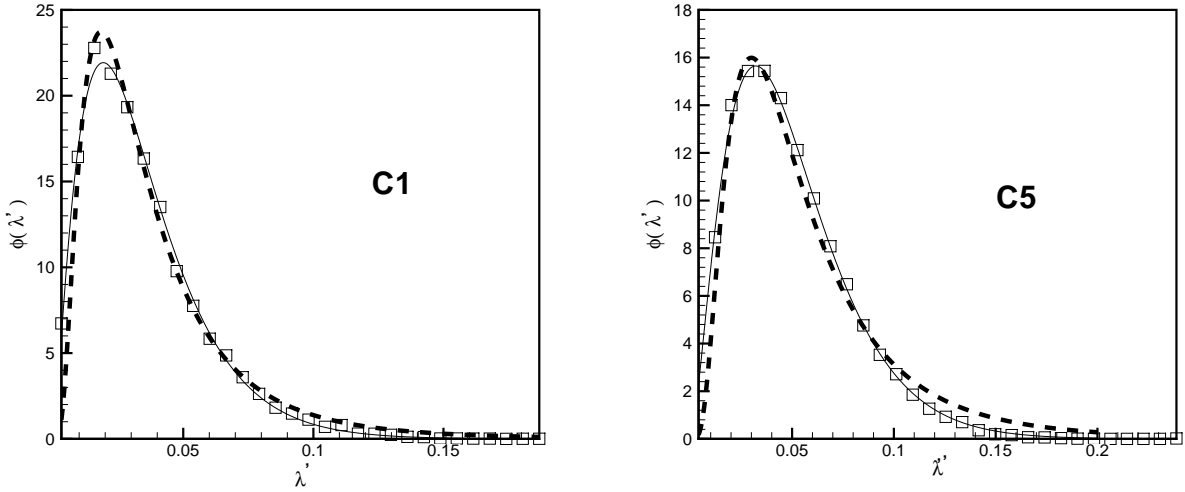}
  \caption{The present-day, at $z=0$, distributions of  $\lambda'$ for the cases C1 and C5 are given by squares at
  the left and the right snapshots, respectively.
  Dashed lines are the best least-squared fits by a log-normal distribution given in text by eq. 14. Solid  lines
  are also the best least-squared by a distribution given by eq. 15. It is clear that the fits by eq. 15 approximate
  better the results than the log-normal distribution. In case C5, this is very clear at the right tail of the
  distribution. }\label{fg3-eps}
    \end{figure*}
    \section{Results}
    We studied five cases for haloes of different present day masses $m_0$. In our system of units, $m_0$ takes the
    values  $ 10^{-2}, 0.1, 1., 10$ and  100 for the cases $1, 2, 3, 4$  and
    $5$, respectively. For every case, we produced a number of $N_{res}=20000-60000$
    realizations. Studying the progenitors of a halo with present day mass $m_0$ at some time $t_*$ it
    is likely to find a number of them with zero angular momenta. This is a consequence of the way of the construction of
    the merger-tree. These progenitors are haloes that have not suffered any merger for times $t\leq t_*$ and thus
    either they have not increase their masses at all or they have accreted only small amounts of matter. Notice that
    the growth of angular momentum resulting by the accretion of amounts of matter that are below
    a critical value -see Hiotelis \& Del Popolo (2006), for the details of the construction of merger trees-
    is not taken into account in our scheme. The distributions presented below are predicted by taking into account
    only haloes that have no zero angular momentum.\\
      We note here that the growth of angular momentum in the above presented picture is not so smooth
    as in the case of the tidal torque theory  where the angular momentum
    is a simple increasing function of time (eg. Barnes \& Efstathiou 1987, Voglis \& Hiotelis 1989). In the picture
    studied in this paper, angular momentum increases or decreases in a complicated
    way. As an example, Fig.2 shows the evolution of $\lambda'$ as
    a function of the scale factor $a$ for randomly selected
    histories from the cases $1,2$ and $3$. This picture is
    characterized by sharp increases and decreases of the spin
    parameter due to mergers. This is a common behavior present
    in similar studies (e.g. Vitvitska et. al 2002).\
    In the following we study four important characteristics, of the above picture of the growth of the angular
    momentum, namely:
    1) The form of the distribution of $\lambda'$.
    2) The dependence of this distribution on the present day mass of the
    halo.
    3) The role of major mergers and their affection on the magnitude and the distribution of $\lambda'$, and
    4) the dependence of the distribution on the redshift.\\

       It has been proposed in the literature that the resulting distribution of $\lambda'$ is
    well approximated by a log-normal distribution with probability density given by

    \begin{equation}
    \phi
    (\lambda')=\frac{1}{\sigma_{\lambda'}\sqrt{2\pi}}exp\left[-\frac{\ln^2(\lambda'/\overline{\lambda'})}
    {2\sigma^2_{\lambda'}}\right]\frac{1}{\lambda'}.
    \end{equation}
    (e.g. van den Bosch 1998; Gardner 2001; Bailin \& Steinmetz
    2005). This is a Gaussian in $\ln\lambda'$. The expectation value for $\ln\lambda'$ is
    $ <\ln\lambda'> =\ln\bar{\lambda}'$ and it peaks at
     $\lambda'_{peak} =\bar{\lambda}'exp(-\sigma^2_{\lambda'})$.
    A recent analysis of large cosmological simulation was performed by Bett et al.2007. These
    authors used the results of the Millenium simulation of Springel et al. (2005), which followed the
    evolution of 10 billion dark matter particles in the
    $\Lambda CDM$ model, to study the properties of more than $10^6$ haloes that were formed.
    Their study showed that the log-normal distribution cannot approximate well the tails of the resulting distribution
    so they proposed other alternatives. An alternative of similar form, but rather more general,
    is used in the present paper. This is of the form:
    \begin{equation}
    \phi(\lambda')=A\left(\frac{\lambda'}{\lambda'_0}\right)^aexp\left[-b\left(\frac{\lambda'}{\lambda'_0}\right)^c\right]
    \end{equation}
    This distribution peaks at ${\lambda'}_{peak}=\lambda'_0(\frac{a}{bc})^{\frac{1}{c}}$
    and the expectation value for $\lambda'$ is given
    by $<\lambda'>=\frac{A{\lambda'}^2_0}{b^{\frac{a+2}{c}}}\Gamma(\frac{a+2}{c})$
    where A, a,b,c,${\lambda'}_0$ are positive parameters and $\Gamma$ is the complete gamma
    function.\
    In Fig.3, we present two characteristic snapshots for the distributions of $\lambda'$, for the cases C1 and C5 that
    correspond to the smallest and the largest haloes studied. Squares correspond to the results predicted by the
    method described in this paper, dashed lines correspond to the best least-squared fits of the results by a log-normal distribution
     while solid lines are the best least-squared  fits by a distribution of the form of eq.15. It is clear that distributions given by
    both eq.14 and eq.15 are good fits of our results for the low
    mass case C1. For case C5, the log-normal distribution cannot
    fit well the tails of the solid line. In both cases, the results are described better by
    eq.15. It is also clear that the distribution depends on the mass of the halo. Although the peaks are
    located at about the same position in both snapshots, the value of the peak is significantly larger in the case
    C1. We also note that $\sigma_{\lambda'}$ is a decreasing function of $m_0$. It varies from 0.71 for case C1 to 0.67 for case  C5.\
    It has also been noticed by other studies (e.g. Vitvitska et al. 2002) that  $<\lambda'>$ does not depend on the mass of the halo.
    Fig.3 reveals that this is not true in our results. This can be seen more clearly in Fig.4, where
     $<\lambda'>$ versus the present-day mass of the halo, $m_0$, is
     plotted. Although for a factor of masses 10000 -from
     0.01 to 100- $<\lambda'>$ varies by a factor only of about 1.41 - from 0.0343 to 0.0484-
     it is obvious that $<\lambda'>$ is an increasing
     function of mass.\
     Checking the ability of formula (15) to fit the results of
     Monte-Carlo predictions we found that for ${\lambda}'_0=0.049$ and $ c=1$ we can predict very good fits to all
     cases, where the values of the rest of the parameters are given in
     Table 1\\
\begin{table}[b]\caption{}
\begin{tabular}{c c c c }
\tableline
       Parameters  & $A$       & $\alpha$ & $\beta$ \\
\tableline
       Case 1      &272.206  & 1.307  & 3.302 \\
       Case 2      &222.880  & 1.292  & 3.032 \\
       Case 3      &183.043  & 1.326  & 2.76  \\
       Case 4      &178.417  & 1.473  & 2.67 \\
       Case 5      &187.216  & 1.719  & 2.66 \\
\tableline
\end{tabular}
\end{table}
     A and $\beta$ seem to be  decreasing functions of mass. As regards $\alpha$, this  is clearly an increasing function
      of the mass of the halo.\\
           According to the results of N-body simulations it is likely that spin parameter is a decreasing
     function of mass. This is supported, for example,  by the results of Bett et al. 2007 and Macci$\grave{o}$ et al. 2007.
     Additionally, Bett et al. showed that the values of spin parameter and its behavior as a function of mass depends
     crucially on the halo-finding algorithm. This conclusion was derived by studying three different halo-finding
     algorithms,  the traditional "friends-of-friends" (FOF) algorithm of Davis et al. (1985), the "spherical overdensity"
     SO) of Lacey \& Cole (1994) and a new halo definition that they introduced, the TREE haloes. Since the results
      are so sensitive to the halo-finding
     algorithm there is a problem regarding the comparison of the results of
     N-body simulations that have used  different halo-finding algorithms. Especially, this problem is more
     serious for the results that are relative to the growth of angular momentum. In any case,  it seems that
     N-body simulations favor a spin parameter that is a decreasing
     function of the virial mass. Instead our results as those of Maller et al. (2002) favor the trend that the
      spin parameter is  an increasing function of the virial mass of the halo. Taking into account that
      our merger-tree algorithm     has been constructed independently of that of the above
      authors, the disagreement with the trend seen in N-body simulations is not likely to arise from
      inconsistencies  in the merger-tree construction algorithm.  Additionally the ellipsoidal collapse model
      we use, that is an improvement of the spherical collapse model used by the above authors, is also
      incapable to resolve the problem. We also have to note here that a similar trend as that seen in our results is
      also found, for $\lambda$,  in the results of Einsenstein \& Loeb (1995) where the collapse of homogeneous ellipsoids
      in the tidal field of their environment is studied. \\
      In order to study the role of the distribution of velocities we also used
      a different distribution according to the predictions of N-body simulations of
      Vitvitska et al 2002 that is defined as follows: Let that haloes with masses $M$ and $m$ and radii $R$ and $r$ merge.
     For convenience capital letters correspond to the larger halo. After the choice $v_{rel}$, as described in section 2.,
     we find $v_{cir,M}=(GM/R)^{1/2}$ and $v_{cir,m}=(Gm/r)^{1/2}$. Then the tangential velocity is picked by a Gaussian
     with mean value $v_{t,mean}=(0.9-0.5\frac{v_{cir,m}}{v_{cir,M}})v_{cir,M}$ for  $v_{cir,m}/v_{cir,M} > 0.4$ and
     $v_{t,mean}=0.7v_{cir,M}$ for $v_{cir,m}/v_{cir,M} < 0.4$. The above Gaussian has $\sigma= v_{t,mean}/3$.
     This scheme is consistent to the results of Colin et al.
     (1999) that major mergers are significantly more radial than minor ones and bring in less specific angular momentum.
      The differences between the results of this new distribution are those derived by the distribution
     described in section 2 are negligible and the spin parameter is again an increasing function of the virial
     mass.  However the question remains open.\\
                A number of authors has stressed the role of recent major merger in the final value of the spin
  parameter. More specifically, they have shown that haloes which have suffered large recent
  major mergers appear to have larger values of spin parameter. We define as recent mergers those occurred at
  redshifts $z\leq 3$. We consider the merge between a large halo $M$ and a small $m$ as major if $m/M \geq 0.5$. Obviously
  a recent major merger satisfies both the above conditions.
  Figs.5 and 6 show the dependence of the $<\lambda'>$ distribution on
  the number of recent major mergers for three of our cases namely C1, C3 and C5. In Fig.6 solid lines show  the distribution of $<\lambda'>$ for all haloes
   of the  respective case, while dashed lines are the distributions over those haloes that had at least
   one recent major merger.  Dotted lines are the distributions of haloes that had no recent major   mergers.
    It is clear, particularly at low masses as in case C1, that dashed lines represent distributions that are shifted to the right
    relative to that represented by solid lines. Thus, haloes that had at least one recent major merger have larger
    spin parameters. On the other hand, haloes that had no recent major merger -represented by dotted lines- show a
    narrow distribution shifted to the left, relative to the solid line, that has an obviously smaller mean value.
  \begin{figure}[b]
    \includegraphics[width=8cm]{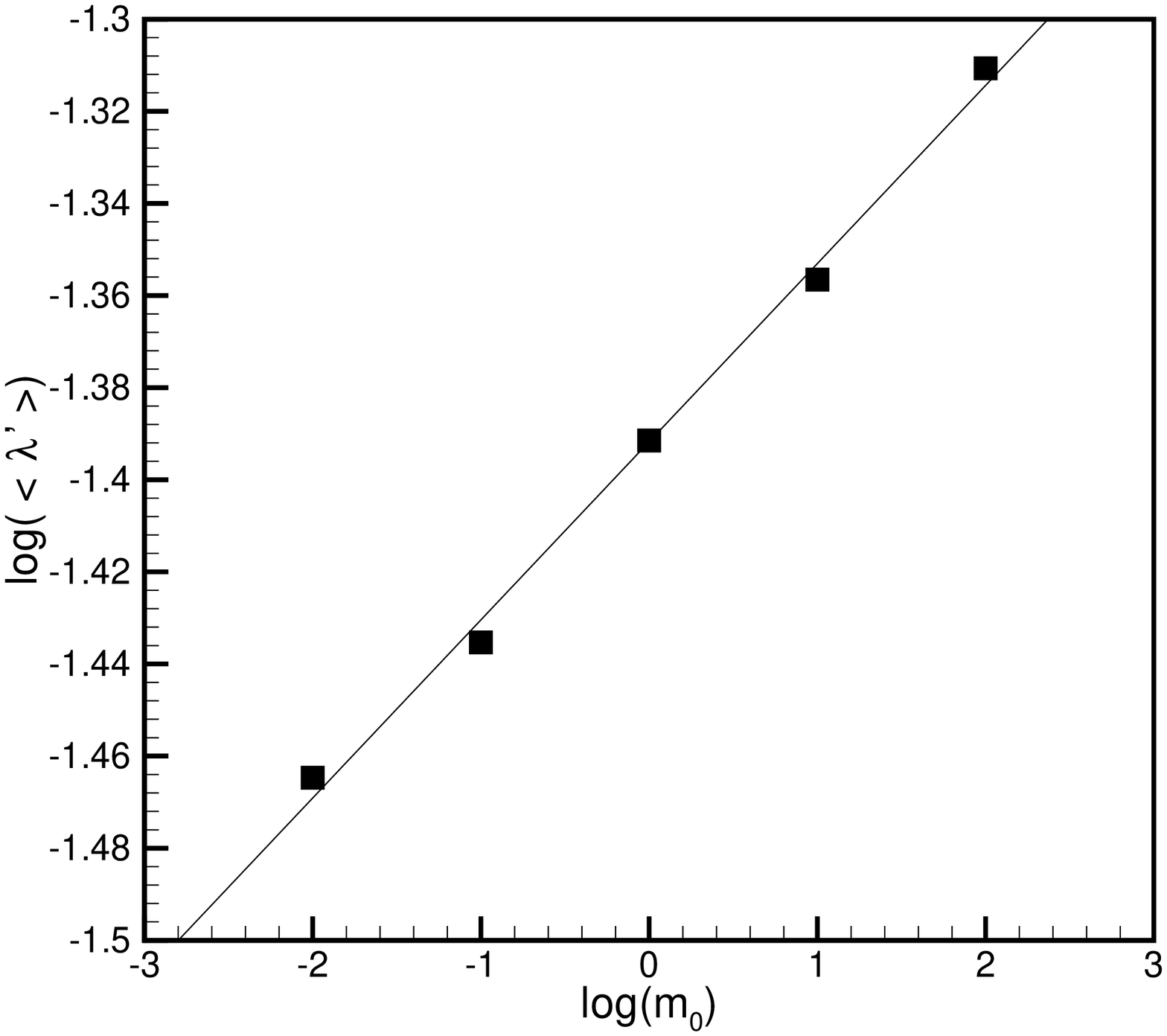}
  \caption{The mean value of $\lambda'$ as a function of the present day mass
  $m_0$ of the halo. Although the values of $\lambda'$ are not very
  sensitive to the values of the mass, as they vary from 0.0343 to 0.0489 for
  a range of mass 0.01 to 100, our results show that $<\lambda'>$ is an increasing function of the
  mass of the halo. The linear dependence shown, $log(<\lambda'>)\approx-1.392+0.0387log(m_0)$,  in this figures indicates that $<\lambda'>$ varies as
  a power of $m_0$.}\label{fg4-eps}
  \end{figure}
    \begin{figure*}[t]
    \includegraphics[width=17cm]{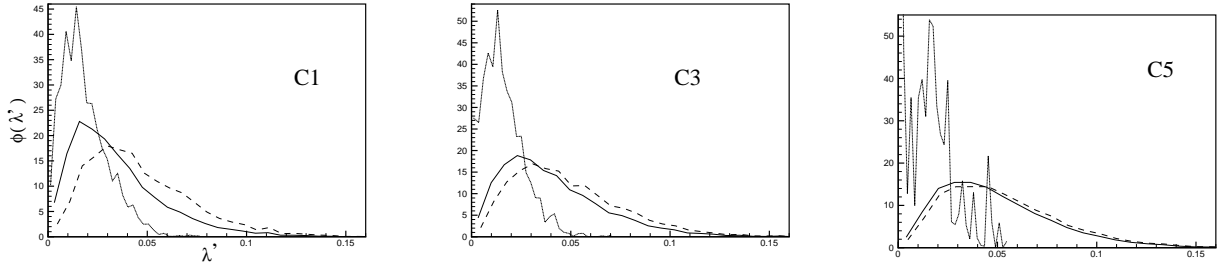}
  \caption{The role of recent major mergers in the distribution of $\lambda'$ for the cases C1,C3 and C5.
  The solid line in every snapshot shows the distribution for all the haloes in every case. Dashed
  lines show the distribution of the haloes that had at least one recent major merger while
  dotted lines correspond to the haloes that had no recent major merger.}\label{fg5-eps}
    \end{figure*}
    \begin{figure}[b]
    \includegraphics[width=8cm]{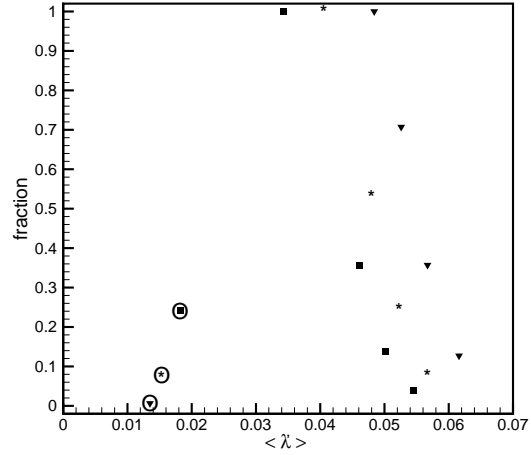}
  \caption{The role of recent major mergers for the cases C1,C3 and C5. The horizontal axis shows values of $<\lambda'>$ while
  the vertical one represent the fractions of various groups of haloes. Squares correspond to the case C1,stars to the case C3 and
   triangles to the case C5. From top to bottom symbols mark the mean value of spin parameter versus
  the fraction for the following groups: i) all haloes of the case,  ii) haloes that had at least one recent major
  merger, iii) haloes that had at least two recent major mergers  and iii),  haloes that had at least three
  recent major mergers. The square in the circle marks $<\lambda'>$ for those haloes of the case C1 that
   had no recent major mergers. The same hold for the star inside the circle and the triangle inside the circle but
   for the cases C3 and C5 respectively.
   See text for more details.}\label{fg6-eps}
   \end{figure}
   \begin{figure}[t]
    \includegraphics[width=8cm]{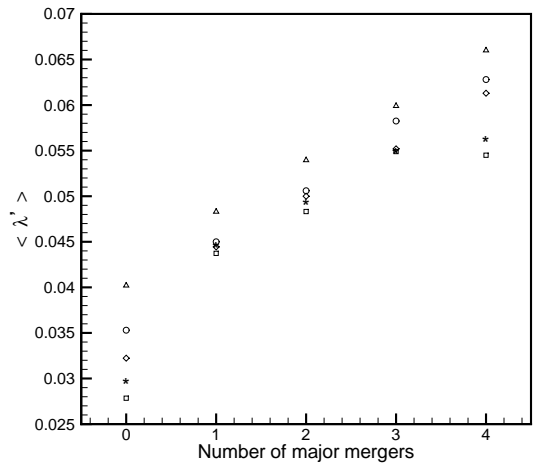}
  \caption{This figure presents the distribution of  spin parameter at fixed number of recent major mergers (0,1,2,3 and 4).
  Different symbols correspond to haloes of different masses. Squares correspond to C1, stars to C2, diamonds to C3,
   circles to C4 and deltas to C5. It is clear that even at fixed number of recent major mergers the distribution of $\lambda'$
   is an increasing function of the mass of the halo. }\label{fg7-eps}
    \end{figure}

    \begin{figure*}[t]
    \includegraphics[width=17cm]{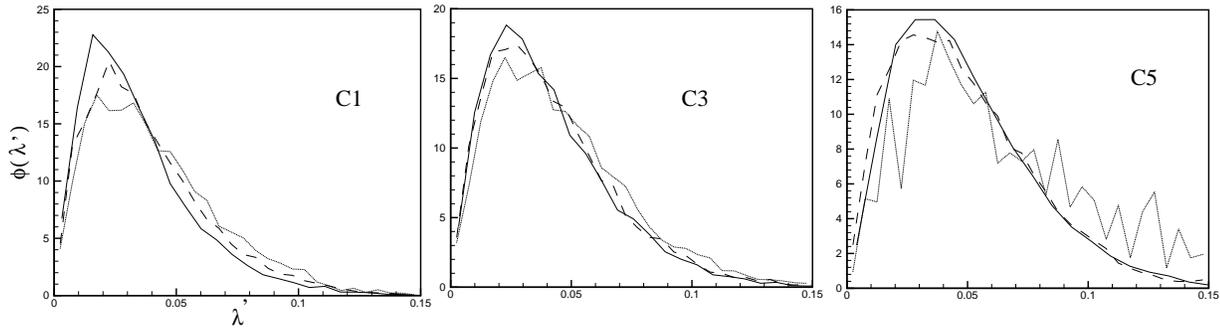}
  \caption{The dependence of the spin distribution on the redshift $z$. Solid lines correspond to present day,
   $z=0$ while dashed and dotted lines correspond to $z=1$ and to $z=3$.
    From the left to the right, snapshots correspond to the cases C1,
   C3 and C5, respectively. It is shown that the distribution of $\lambda'$ is, to a good approximation, unchanged
   between $z=0$ and $z=1$. Differences between $z=0$ and $z=3$ are more obvious but without a significant change
   of the shape of the distribution. }\label{fg8-eps}
    \end{figure*}

  In Fig. 6 we present $<\lambda'>$ for various groups of haloes from the cases C1 and C5 versus the
  fraction of the total number of haloes of the case that these groups represent. Squares correspond to the case C1, stars
  to the case C3 and triangles to the case C5. From the left to the right symbols mark the mean value of spin parameter versus
  the fraction for the following groups: i) all haloes of the case,  ii) haloes that had at least one recent major
  merger, iii) haloes that had at least two recent major mergers  and iv) haloes that had at least three
  recent major mergers. The square inside the circle marks $<\lambda'>$ for those haloes of the case C1 that
  had no recent major mergers. The same is indicated by the star inside the circle and the triangle inside the circle for
  the cases C3 and C5, respectively. This fig. shows that $<\lambda'>$ is an increasing function of the number of recent major mergers. For example, the
  group, from case C1, of haloes that had at least one recent major merger has $<\lambda'>= 0.0461$ and represents
  $ 35.6 $\% of the total haloes of the case while the group of haloes that had at least three major merger has $<\lambda'>=
  0.0545$ and represents only $4$\% of the total number of haloes of the
  case. For the group of haloes that had no recent major merger $<\lambda'>=0.0182$, a significantly low value,
  while the
  haloes of this group represent $24.3$\% of the total number of haloes of the
  case. It is noticed that as the haloes belonging to that last group had an unperturbed recent history, they have
  time to evolve their gas smoothly to a rotationally supporting disk. Its fraction is an decreasing function of
  the final halo mass. For case  C5 only $0.63$\% belong to that group. The fraction of
  haloes with no recent major merger is definitely a decreasing function of
  the present day mass of the halo. It is quite reasonable, in the hierarchical clustering scenario studied here,
  that recent major mergers become rare effects for small
  haloes.\
  It would be interesting to see if, at fixed number of recent major
  mergers, the distribution of $\lambda'$ is independent of mass.
  For this reason, in Fig.7 we plotted the mean value of $\lambda'$ for haloes
  that have suffered the same number of recent major
  mergers(0,1,2,3 and 4) for all cases. Different symbols correspond to haloes of different masses.
  Squares correspond to C1, stars to C2, diamonds to C3,
   circles to C4 and deltas to C5.  It is clear from this fig. that for haloes
  that have suffered the same number of recent major mergers, the heavier
  one has the larger $\lambda'$.\
    Summarizing the results it is yielded that:
  a) In all cases haloes that had at least one recent major merger have
  $<\lambda'>$ larger than those haloes that had none.
  b) The fraction of haloes that had at least one recent major merger is larger than
  the fraction of haloes that had no recent major mergers at all.
    Thus, we can draw the following conclusions:\\
    1) Disk galaxies are found preferentially in small haloes.\\
    2) Between haloes of the same mass those that host
  elliptical galaxies rotate faster than haloes of spiral
  galaxies (see Vitvitska et al. 2002).\\
    3) The number of haloes that had no recent major merges is significantly smaller that the number of haloes
    that had at least one major merger. If we took into account that major mergers destroy galactic disks and produce
    spheroidal stellar systems (see, e.g., Barnes 1999)  small mergers probably do not
    (see, e.g., Walker, Mihos \$ Hernquist 1996) it is natural to
    expect that haloes which have suffered at least one major
    merger could not a host a spiral galaxy. However spheroidal
    stellar systems should be more common objects than spiral
    galaxies.\\
  Fig.8 depicts the dependence of the distribution of spin parameter on the redshift
  $z$. Solid lines  correspond to  present day $z=0$ while dashed  and dotted lines to $z=1$ and to $z=3$ respectively.
  From the left to the right, snapshots correspond to the cases C1,
   C3 and C5 respectively.  Differences between $z=0$ and $z=1$ are small. Curves at  $z=1$ appear with smaller peaks
  and a slight shift, relative to curve for $z=0$, to the right for small haloes and to the left for larger haloes.
  Differences between $z=0$ and $z=3$ are more obvious. Dashed  curves are, for all cases, shifted to the right relative
  to solid  curves. The form of the distributions remains practically unchanged. A shift of the distribution
  to  the right shows that the mean value of the spin parameter becomes larger while a shift to the left  shows that
  it becomes smaller. Thus, the curves  indicate that the value of the spin parameter decreases from $z=3$ to
  $z=0$ for all cases. This is verified by the straightforward evaluation of $<\lambda'>$ that appears to be systematically
  smaller at $z=0$ than its value at $z=3$.

\section{Discussion}

  This study describes a picture for the growth of the angular momentum of dark matter
  haloes in terms of a hierarchical clustering scenario. The results presented above are, in general, in good agreement with
  the results that were already known in the literature. Comparing our results with those of large N-body
  simulations,
  we have found satisfactory agreement in the following points:\\
  1) The values of  spin parameter are in the range of  0.0343 to 0.0484 for haloes with present
      day masses in the range of $ 10^9\mathrm{h}^{-1}M _{\odot}$ to
      $10^{14}\mathrm{h}^{-1}M_{\odot}$.\\
  2) A log-normal distribution approximates satisfactorily the distributions of the values of the spin parameter but
      it fails to describe accurately the tails of the resulting distributions. A new, more satisfactory formula,
      is presented.\\
  3) The role of recent major mergers is very important. The distribution  of the spin parameter is appreciably affected by the
     number of recent major merger. The present day value of the spin parameter of a halo is an increasing
     function of the number of the recent major mergers.\\
  4) The distributions  of the spin parameter do not depend significantly on the redshift.\\
  5) The value of the spin parameter is a function of the present day mass of the halo. The form of this
     function depends, in N-body simulations, on the halo-finding algorithm but in general seems that spin parameter is
     a decreasing function of mass. Instead, in our results, $<\lambda'>$ is an
     increasing function of mass, approximately very closely a power-law form.\\
    Our results give rise to some  questions, as for example:  Why semi-analytical methods
     are not able to predict the correct relation between the spin parameter
  and the virial mass of the halo?  Does this disagreement reflects the null role of tidal fields in the
  orbital-merger picture or it arises from other problems associated with the nature of merger-trees?  Are merger-trees able
  to give the correct relation for better description of the velocity field during the merge?
  Is any way improvements on both
  analytical  and numerical methods are required in order to help us answering  some of the above
   questions and to advance our understanding about the
   physical processes that created the structures we observe.
\section{Acknowledgements}  We acknowledge the anonymous referee for useful comments and suggestions, Dr. M.
Vlachogiannis and K. Konte for assistance in manuscript
preparation, and the \textit{Empirikion} Foundation for its
financial support.


\end{document}